\documentclass[a4paper,UKenglish,cleveref, autoref, thm-restate]{lipics-v2021}

\usepackage{thm-restate}

\usepackage{scalerel}

\newcommand{\bigRectangle}{{
    \ooalign{$\sqsubset\mkern2mu$\cr$\mkern2mu\sqsupset$\cr}
}}

\newcommand{\rect}{{
\scaleto{\bigRectangle}{4pt}}}

\newcommand{\ceil}[1]{\left\lceil{#1}\right\rceil}
\newcommand{\floor}[1]{\left\lfloor{#1}\right\rfloor}

\newcommand{\R}{\mathbb{R}}
\newcommand{\N}{\mathbb{N}}
\renewcommand{\P}{\mathcal{P}}

\newcommand{\maximum}{\mathsf{Maximum}}
\newcommand{\minimum}{\mathsf{Minimum}}
\newcommand{\argmax}{\mathsf{argmax}}
\newcommand{\argmin}{\mathsf{argmin}}
\newcommand{\poly}{{\mathsf{poly}}}
\newcommand{\polylog}{{\mathsf{polylog}}}

\newcommand{\Occ}{\mathsf{Occ}}
\newcommand{\SA}{\mathsf{SA}}
\newcommand{\LSA}{\mathsf{LSA}}
\newcommand{\ILSA}{\mathsf{ILSA}}
\newcommand{\LCP}{\mathsf{LCP}}
\newcommand{\SCDS}{\mathsf{SCDS}}
\newcommand{\cI}{\mathcal{I}}
\newcommand{\M}{\mathcal{M}}

\newcommand{\per}{\mathsf{per}}

\newcommand{\anull}{\mathsf{null}}

\renewcommand{\S}{\mathcal{S}}

\renewcommand{\int}{\mathsf{Int}}
\newcommand{\modulo}{\operatorname{mod}}

\newcommand{\MaxComp}{\mathsf{MaxCompatible}}
\newcommand{\point}{\mathsf{point}}
\newcommand{\rows}{\mathsf{rows}}
\newcommand{\columns}{\mathsf{columns}}

\newcommand{\eps}{\varepsilon}

\newcommand{\lef}{\mathsf{left}}
\newcommand{\righ}{\mathsf{right}}

\newcommand{\Otild}{\tilde{O}}

\crefname{enumi}{Property}{Properties}

\newcommand{\para}[1]{\subparagraph*{#1}}

\newtheorem{fact}[theorem]{Fact}

\title{Searching 2D-Strings for Matching Frames} 
\author{Itai Boneh}{Reichman University and University of Haifa, Israel}{itai.bone@biu.ac.il}{https://orcid.org/0009-0007-8895-4069}{supported by Israel Science Foundation grant 810/21.}

 \author{Dvir Fried}{Bar Ilan University, Israel}{friedvir1@gmail.com}{https://orcid.org/0000-0003-1859-8082}{supported by ISF grant no. 1926/19, by a BSF grant 2018364, and by an ERC grant MPM under the EU's Horizon 2020 Research and Innovation Programme (grant no. 683064).}

  \author{Shay Golan}{Reichman University and University of Haifa, Israel}{golansh1@biu.ac.il}{https://orcid.org/0000-0001-8357-2802}{supported by Israel Science Foundation grant 810/21.}

    \author{Matan Kraus}{Bar Ilan Univesity, Israel}{matan3@gmail.com}{https://orcid.org/0000-0002-2989-1113}{supported by the ISF grant no. 1926/19, by the BSF grant 2018364, and by the ERC grant MPM under the EU's Horizon 2020 Research and Innovation Programme (grant no. 683064).}
        
    \author{Adrian Micl\u au\c s}{ Faculty of Mathematics and Computer Science, University of Bucharest, Romania}{adrian.miclaus1@gmail.com}{https://orcid.org/0009-0001-6988-185X}{supported by a grant of the Ministry of Research, Innovation and Digitization, CNCS - UEFISCDI, project number PN-III-P1-1.1-TE-2021-0253, within PNCDI III.}

    \author{Arseny Shur}{Bar Ilan University, Israel}{shur@datalab.cs.biu.ac.il}{https://orcid.org/0000-0002-7812-3399}{supported by the ERC grant MPM under the EU's Horizon 2020 Research and Innovation Programme (grant no. 683064) and by the State of Israel through the Center for Absorption in Science of the Ministry of Aliyah and Immigration.}

\authorrunning{I. Boneh et al.} 

\Copyright{Itai Boneh, Dvir Fried, Shay Golan, Matan Kraus, Adrian Micl\u au\c s,  and Arseny Shur} 


\nolinenumbers 
\hideLIPIcs
\begin{document}

\maketitle

\begin{abstract}
We study a natural type of repetitions in $2$-dimensional strings.
Such a repetition, called a matching frame, is a rectangular substring of size at least $2\times 2$ with equal marginal rows and equal marginal columns.
Matching frames first appeared in literature in the context of Wang tiles.

We present two algorithms finding a matching frame with the maximum perimeter in a given $n\times m$ input string.
The first algorithm solves the problem exactly in $\Otild(n^{2.5})$ time (assuming $n \ge m)$.
The second algorithm finds a $(1-\eps)$-approximate solution in $\Otild(\frac{nm}{\eps^4})$ time, which is near linear in the size of the input for constant $\eps$. 
In particular, by setting $\eps = O(1)$ the second algorithm decides the existence of a matching frame in a given string in $\Otild(nm)$ time.
Some technical elements and structural properties used in these algorithms can be of independent interest.
\end{abstract}
\clearpage
\setcounter{page}{1} 
\section{Introduction}
\label{sec:intro}

Throughout the years, a variety of notions for repetitive structures in strings have been explored; see, e.g.,~\cite{FW65,Manacher75,KMP77,Weiner73,KK99}.
Even recently, new efficient algorithms regarding palindromes~\cite{BKRS17,GMSU19,RuSh20}, squares~\cite{EGG23}, runs~\cite{BIINTT17,EF21,MeSh19}, and powers~\cite{ABCK19} have been introduced. 
In the studies on $2$-dimensional strings (aka \emph{$2$d-strings} or \emph{matrices}), periodic and palindromic structures also attracted definite interest \cite{AG92,AG98,ALMS2020,CRRWZ20,GGL21,KM17,GRSS17,Smith17}.

\begin{figure}[ht!]
  \begin{center}
 \includegraphics[scale=0.5]{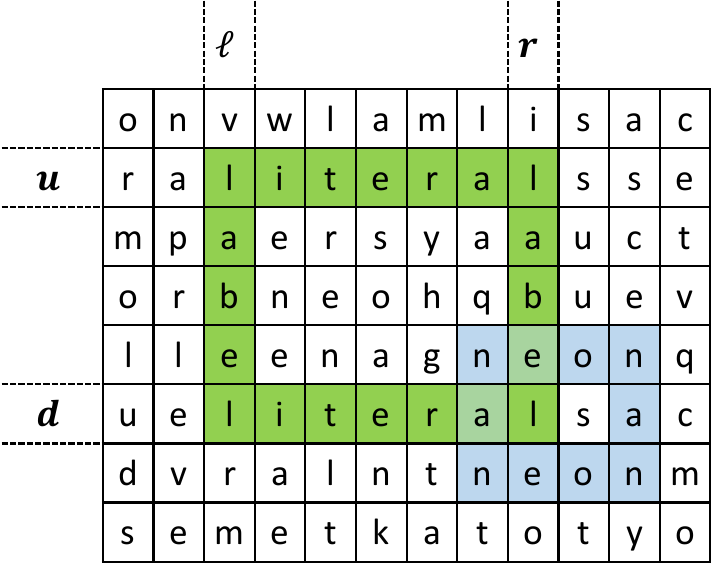} 
  \caption{An example of a matching frame $(u,d,\ell,r)=(2,6,3,9)$. 
  The strings on the top and bottom sides of the frame are equal, and the strings on the left and right sides are also equal. 
  The perimeter of the frame is $2\cdot(6-2+9-3)=20$.
  The matrix also contains a smaller matching frame.
  \label{fig:udlrexample}}
   \end{center}
   \vspace*{-2mm}
\end{figure}

Matching frame is a natural repetition in $2$d-strings, first considered by Wang \cite{Wang61} when introducing Wang tiles.
Given a $2$d-string $M$ over an alphabet $\Sigma$, 
a \emph{frame} in $M$ is a rectangle defined by a tuple $(u,d,\ell,r)$ such that $u<d$ and $\ell<r$. 
This rectangle covers the submatrix $M[u..d][\ell..r]$ and is \emph{matching} if this submatrix has equal marginal rows and equal marginal columns. 
Formally, $(u,d,\ell,r)$ is a matching frame if $M[u][\ell.. r] = M[d][\ell.. r]$ and $M[u.. d][\ell] = M[u..d][r]$ (see \cref{fig:udlrexample}).
Wang's \emph{fundamental conjecture}, later disproved by Berger \cite{Ber66}, said ``a set of tiles is solvable (= tiles the plane) if and only if it admits a cyclic rectangle (= matching frame)''.
Note that a fast algorithm to find matching frames would simplify a huge computation conducted by Jeandel and Rao \cite{JeRa15} to prove that their aperiodic set of tiles is minimal.

Matching frames indicate ``potential'' periodicity in two dimensions. 
Namely, if a $2$d-string $M$ is built according to some local rule, then any matching frame in $M$ can be extended to a periodic tiling of the plane, \emph{respecting this local rule}. 
Well-known examples of such local rules are given, in particular, by self-assembly models such as aTAM \cite{RoWi00} or 2HAM \cite{CDDEPSSW13}.
Note that matching frame is an \emph{avoidable} repetition: as was first observed by Wang \cite{Wang65}, there exist infinite \emph{binary} $2$d-strings without matching frames. 
Avoidable repetitions are interesting, in particular, due to a nontrivial decision problem.

Overall, there is a clear motivation to design efficient algorithms searching for matching frames.
Let us specify the exact problem studied in this paper.
The \emph{perimeter} of a frame $F = (u,d,\ell,r)$ is the total number of cells in its marginal rows and columns, i.e. $\per(F) = 2(d-u+r-\ell)$. 
By \emph{maximum} frame (in a set of frames) we mean the frame with the maximal perimeter in this set.
In the \textit{maximum matching frame problem}, the goal is to find a maximum matching frame in a given matrix or report that no matching frame exists. 
We also consider the $(1-\eps)$-approximation version of this problem, in which the goal is to find a matching frame with a perimeter within the factor $(1-\eps)$ from the maximum possible.

\para{Our Results.} We present $\Otild(nm)$-space algorithms that establish the following bounds on the complexity of the  maximum matching frame problem and its approximation version.
\begin{theorem}[Maximum Matching Frame]\label{thm:max_frame}
    The time complexity of the maximum matching frame problem for an $n\times m$ matrix $M$ is $\Otild(n^{2.5})$ in the case $m=\Theta(n)$. 
    In the general case, the complexity is $\Otild(ab\min\{a,\sqrt{b}\})$, where $a=\min\{n,m\}$ and $b=\max\{n,m\}$.\footnote{Throughout the paper, $\Otild(f(n)) = O(f(n) \cdot \polylog n)$}
\end{theorem}

\begin{theorem}[($1-\eps$)-Approximation] \label{thm:apx}
    The time complexity of the $(1-\eps)$-approximation maximum matching frame problem for an $n\times m$ matrix $M$ is $\Otild(\frac{nm}{\eps^4})$.
\end{theorem}

\begin{corollary}[Deciding Matching Frame]
    There is an algorithm deciding whether an $n\times m$ matrix $M$ contains a matching frame in $\Otild(nm)$ time and space.
\end{corollary}

We remark that our exact and approximation algorithms can be straightforwardly adapted to find matching frames with the maximum area / the minimum perimeter / the minimum area instead of matching frames with the maximum perimeter.

\subsection{High-Level Overview}
\para{Maximum Matching Frame.}
The algorithm for finding a maximum matching frame follows a heavy-light approach.
The parameter used to distinguish between heavy and light frames is the \textit{shorter side} of the frame.
A frame $F=(u,d,\ell,r)$ has \emph{height} $d-u$ and \emph{width} $r - \ell$. 
We assume that there is a maximum matching frame having its height smaller than or equal to its width.
(Either the input matrix or its transpose satisfies this assumption and we can apply our algorithm to both matrices and return the better of two results.) 
For some integer threshold $x$, we say that a frame with $d-u \le x$ is \textit{short} (or light); otherwise, it is \textit{tall} (or heavy).
We provide two algorithms,
one that returns a maximum \emph{short} matching frame in $M$ and another returns a maximum \emph{tall} matching frame in $M$.
The largest of the two answers is the maximum matching frame in $M$.

The algorithm for short frames iterates over all pairs of rows with distance at most $x$ from each other.
Note that there are $O(nx)$ such pairs.
Moreover, under the assumption that some matching frame $F=(u,d,\ell,r)$ is short, the rows $u$ and $d$ used by $F$ are processed as a pair.
When processing a pair, the algorithm decomposes its rows into maximal equal segments.
Every segment is processed in linear time to obtain a maximum matching frame that uses a portion of the segment as top and bottom rows (see \cref{subsec:short}).
The accumulated size of the segments is bounded by $m$, so the algorithm runs in $\Otild(n \cdot m \cdot x)$ time.

The algorithm for tall frames (see \cref{subsec:tall}) first guesses a range $[H/2 ..H]$ for the height and a range $[W/2.. W]$ for the width of a maximum matching frame. 
As we consider tall frames, the ranges are sufficiently large, so it is easy to find a small set of positions $\P$ in the matrix $M$ such that every frame with the height and width from the given ranges contains a position from $\P$.
The algorithm employs a subroutine that, given $H,W$, and a position $(i,j)$, computes a maximum matching frame among the frames that contain $(i,j)$, have the height in $[H/2..H]$ and the width in $[W/2..W]$.
The implementation of this subroutine is the main technical part of the algorithm.
This is done by maintaining and querying a range data structure  (see \cref{sec:SCDS}) that allows one to process pairs of columns and pairs of rows with the position $(i,j)$ between them.
There are $O(W^2)$ pairs of columns and  $O(H^2)$ pairs of rows to be processed, which we do in $\Otild(H^2+W^2) = \Otild(W^2)$ total time.
We also show that $|\P| = O(\frac{nm}{HW})$, and therefore the running time for one pair of ranges is $\Otild(nm\frac{W}{H})$.
We further observe that the sum of values $\frac{W}{H}$ over all guessed ranges is $O(\frac{W'}{H'})$ for some single guessed pair $(W',H')$.
Since $x\le H'\le W'\le \max\{n,m\}$, we obtain the running time of $\Otild(nm\frac{\max(n,m)}{x})$.

Finally, the algorithm selects the threshold $x = \sqrt{\max\{n,m\}}$ and applies the algorithms for both the short and the tall case to obtain a running time of $\Otild(nm\sqrt{\max\{n,m\}})$.
Alternatively, one can run the algorithm for short frames alone, setting $x=\min(n,m)$.
Taking the better of these two options proves \cref{thm:max_frame}.

\para{Approximation Algorithm.}
As a preliminary step in our approach for finding a $(1-\eps)$-approximation to the maximum matching frame, we apply a two-dimensional variant of the so-called \emph{standard trick}~\cite{CC07,CKRPRWZ22} 
from certain one-dimensional pattern matching problems.
In pattern matching, we are given a text $T[1..n]$ and a pattern $P[1..m]$ and the goal is to find all the indices $i \in [n-m+1]$ such that $T[i.. i{+} m {-}1]$ ``matches'' $P$.
The standard trick refers to partitioning $T$ into $O(n / m)$ overlapping fragments of size $\Theta(m)$, such that every match of $P$ is contained in a fragment. 
In general, the trick allows one to assume that the length of the text is within a small factor from the length of the pattern. 
Our two-dimensional variant of this trick (\cref{lem:reduction}) allows us to assume that both dimensions of the maximum matching frame are within a $\poly(1-\eps)$ factor of the vertical and the horizontal lengths of $M$.

This assumption allows us to focus on matching frames with sides that are ``close'' to the boundaries of $M$; we call such frames \emph{large}.
The algorithm uses a carefully selected threshold for being close to the boundaries, guaranteeing that
(1) the maximum matching frame is large and 
(2) the perimeter of every large frame approximates the  perimeter of the maximum matching frame.
With that, the problem boils down to determine whether there exists a large matching frame.
The main technical novelty of the approximation algorithm is solving this decision problem in near-linear time.

The algorithm for the above decision problem consists of two main components. 
The first component (see \cref{sec:app_dec}) is an $\Otild(1)$ time subroutine that, given a triplet $(u,d,\ell)$, decides if there is an integer $r$ such that $(u,d,\ell,r)$ is a large matching frame.
However, applying this subroutine to every triplet would cost $\Omega(n^2m)$ time.
The second component (see \cref{sec:interesting_algo}) of the algorithm is the retrieval of a set of $\Otild(nm)$ triplets such that if some large matching frame exists, there must also be a large matching frame derived from one of these triplets.

We conclude by presenting the combinatorial structure that allows us to consider $\Otild(nm)$ triplets in the second component. 
Consider a triplet $(u,d,\ell)$ and let $k$ be the largest integer such that $M[u][\ell .. k]=M[d][\ell .. k]$ (let $S$ denote this string).
Assuming there exists an index $r$ such that $(u,d,\ell,r)$ is a large matching frame, one has $r\le k$.
Observe that if there is an index $d' < d$ that is close to the bottom boundary of $M$ such that $M[d'][\ell..k]=S$, then $(u,d',\ell,r)$ is also a large matching frame. 
Therefore, the triplet $(u,d,\ell)$ can be removed from the set of triplets that have to be processed.
We say that a triplet that is not eliminated due to this reasoning is \emph{interesting}.
Surprisingly, the number of interesting triplets is bounded by $O(nm\log n)$ (see \cref{sec:interesting_combi}).
This combinatorial observation is the main novelty of the approximation algorithm.

\section{Preliminaries}\label{sec:preliminaries}

We use range notation for integers and strings.
We write $[i.. j]$ and $[i.. j)$ for the sets $\{i,\ldots,j\}$ and $\{i,\ldots,j-1\}$ respectively (assuming $i\le j$).
Further, we abbreviate $[1..n]$ to $[n]$.
A string $S[1.. n]=S[1]S[2]\cdots S[n]$ is a sequence of characters from an alphabet $\Sigma$.
We also write $\overleftarrow{S[1.. n]}=S[n]S[n{-}1]\cdots S[1]$.
For every $i\le j\in[n]$, $S[i .. j]=S[i]S[i+1]\cdots S[j]$ is a \emph{substring} of $S$.
The substring is called a \emph{prefix} (resp., a \emph{suffix}) of $S$ if $i=1$ (resp.,  $j=n$).
We assume $\Sigma$ to be linearly ordered, inducing a \emph{lexicographic order} (\emph{lex-order}) on strings. 

An $n\times m$ \emph{matrix} (or \emph{$2$d-string}) $M$ is a 2-dimensional array of symbols from $\Sigma$.
We refer to the number of cells in $M$ as the \textit{size} of $M$, writing  $|M| = nm$.
We denote a horizontal substring of $M$ as $M[i][j_1 .. j_2] = M[i][j_1]M[i][j_1 + 1] \ldots M[i][j_2]$. 
Similarly, we denote a vertical substring as $M[i_1 .. i_2][j]= M[i_1][j]M[i_1+1][j] \ldots M[i_2][j]$.

\subsection{Suffix Arrays, Longest Common Prefixes}
For a tuple of strings $\S=(S_1,S_2,\dots,S_n)$, the \emph{lexicographically sorted array} $\LSA_\S$ is an array of length $n$ that stores the lex-order of the strings in $\S$.
Formally, $\LSA_\S[i]=j$ if $S_j$ is the $i$th string in $\S$ according to the lex-order (ties are broken arbitrarily).
For a string $S[1.. n]$, the \emph{suffix array} $\SA_S$ of $S$ is the $\LSA$ of all suffixes of $S$.
Formally, for every $i\in[n]$ let $S_i=S[i..n]$ and let $\S_S=(S_1,S_2,\dots,S_n)$; then $\SA_S=\LSA_{\S_S}$.

The suffix arrays were introduced by Manber and Myers \cite{MaMy90} and became ubiquitous in string algorithms. The array can be constructed in near-linear time and space by many algorithms  \cite{KS03,KSPP03,KA05,NZC09dcc,NZC09cpm,Weiner73,Ukkonen95}. 

\begin{lemma}\label{lem:sa_construct}
Given a string $S[1..n]$, the suffix array of $S$ can be constructed in $O(n\log n)$ time and space.
\end{lemma}

An important computational primitive is a data structure for computing the length of the \emph{longest common prefix} of two strings $S[1..n]$ and $T[1..m]$, given as $\LCP(S,T)=\max\{\ell\in[\min\{n,m\}] \mid S[1..\ell]=T[1..\ell]\}$.
An $\LCP$ data structure $\LCP_\S$ for a set of strings $\S=\{S_1,S_2,\dots,S_n\}$ supports queries in the form ``given two indices $i,j\in[n]$, report $\LCP(S_i,S_j)$''.
We denote by $\LCP(S)$ the $\LCP$ data structure for the set of suffixes of a given string $S[1..n]$.
It is known that the following can be obtained by applying the lowest common ancestor data structure of \cite{HT84} to the suffix tree of \cite{Weiner73}.

\begin{lemma}\label{lem:LCPds} 
There is an $\LCP$ data structure with $O(n\log n)$ construction time and $O(1)$ query time. The data structure uses $O(n)$ space. 
\end{lemma}

The following facts are easy.
We give their proofs for the sake of completeness.

\begin{fact}
\label{fact:3LCP}
    Given three strings $S_1,S_2$ and $S_3$, the condition $\LCP(S_1,S_2)> \LCP(S_1,S_3)$ implies $\LCP(S_1,S_3)=\LCP(S_2,S_3)$.
\end{fact}    
\begin{proof}
    Let $\LCP(S_1,S_3)=k$.
    One has $S_1[1..k]=S_2[1..k]=S_3[1..k]$ and $S_1[k+1]=S_2[k+1]\ne S_3[k+1]$.
    Then $\LCP(S_2,S_3)=k$ by definition.
\end{proof}

\begin{fact}
\label{lem:sa_fingerprint}
Let $\S = (S_1,S_2, \ldots , S_n)$ be a tuple of strings and let $P[1..m]$ be a string.
The set $\Occ(\S,P)=\{k \mid S_k[1.. m]  = P\}$ coincides with the range $\LSA_\S[i.. j]$ for some $i,j \in [n]$. 

Furthermore, there is an $O(\log n)$ time algorithm that given $k$, $m$, $\LSA_\S$, and $\LCP_\S$ computes $i$ and $j$ such that $\Occ(\S,S_k[1..m]) = \LSA_\S[i..j]$.
\end{fact}

\begin{proof}
    For the first statement it suffices to note that, in a lex-sorted list, all strings $S_k$ with $S_k[1..m]=P$ follow all strings $S_x$ with $S_x[1..m]<P$ and precede all strings $S_y$ with $S_y[1..m]>P$.
    
    The required algorithm can be obtained in two steps as follows.
    First, apply a binary search on $\LSA_\S$ for an index $k'$ that satisfies $\LSA_\S[k']=k$ (Note that the lex-order between two strings in $\S$ can be decided using an $\LCP$ query). 
    Then, apply a binary search in both directions of $k'$ to find the minimal and maximal indices $i \in [1 .. k']$ and $j \in [k' .. n]$ such that $S_{\LSA[i]}[1..m] = S_k[1..m]$ and $S_{\LSA[j]}[1..m] = S_k[1..m]$.    
\end{proof}

\begin{definition}[Fingerprint]\label{def:fingerprint}
For a tuple $\S$ and a string $P=S_k[1..m]$, the \emph{fingerprint} of $P$ in $\S$ is the tuple $(i,j,m)$ such that $i$ and $j$ are the indices specified in \cref{lem:sa_fingerprint}.     
\end{definition}

\subsection{Orthogonal Range Queries}
Our algorithms use data structures for \emph{orthogonal range queries}.
Such a data structure stores, for some positive integer dimension $d$, a set $\P \subseteq \R^d$ of $d$-dimensional points.
Each point $p\in\P$ has an associated value $v(p) \in \R$.
The data structure supports the queries regarding an input $d$-dimensional orthogonal range $R = [a_1 .. b_1]\times [a_2 .. b_2] \times  \ldots  \times [a_d .. b_d]$.
For a point $p = (x_1,x_2, \ldots, x_d)$ one has $p\in R$ if $x_i \in [a_i .. b_i]$ for every $i\in [1.. d]$.
We need the queries $\maximum(R) = \argmax_{v(p)} (p \in R \cap \P) $ and $\minimum(R)=\argmin_{v(p)} (p \in R \cap \P)$.
For this, we use the data structure~\cite{Willard85,Chazelle88} with the following running times.

\begin{lemma}\label{lem:RangeQueries}
For any integer $d$, a set of $n$ points in $\R^d$ can be preprocessed in $O(n\log^{d-1}n)$ time and space to support $\maximum$ and $\minimum$ range queries in $O(\log^{d-1} n)$ time. 
\end{lemma}

In \cref{sec:app_dec}, we use a very particular type of 2-dimensional $\maximum$/$\minimum$ queries, where $v(p)$ is one of the coordinates of $p$. 
Though faster data structures are known in this case \cite{BePu16, GHN20}, using these data structures cannot improve the asymptotics of our results.

\section{Data Structures}\label{sec:preproc}
\begin{figure}[ht!]
  \begin{center}
    \begin{subfigure}[r]{\textwidth}
  \begin{center}
   \caption{}
 
 \includegraphics[scale=0.6]{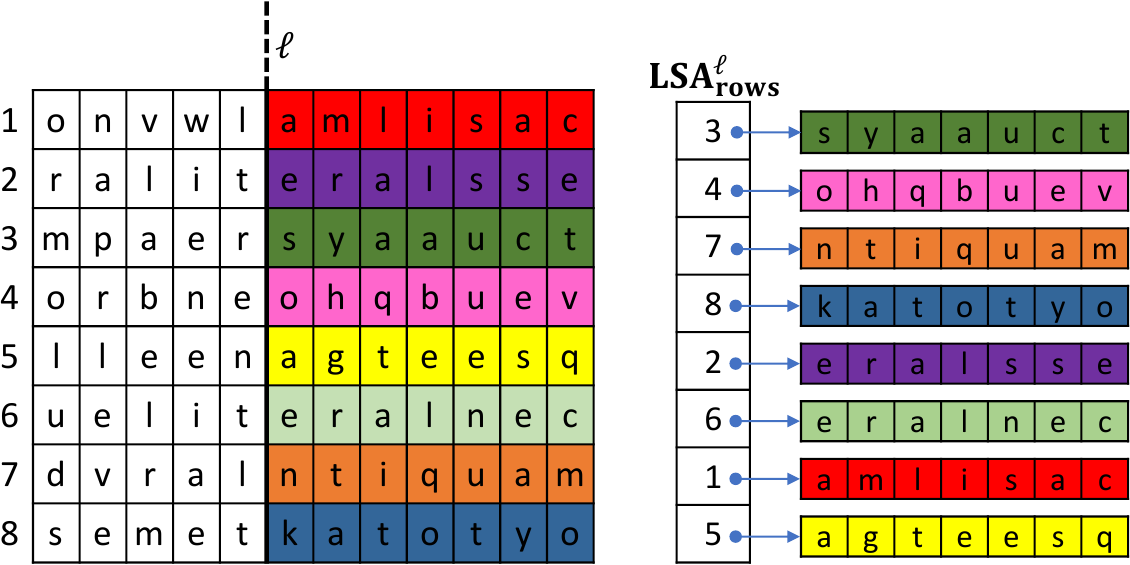} 
    \label{fig:subA}
    \end{center}
 \end{subfigure}\\

 \begin{subfigure}[r]{\textwidth}
   \begin{center}
   \caption{}

 \includegraphics[scale=0.6] {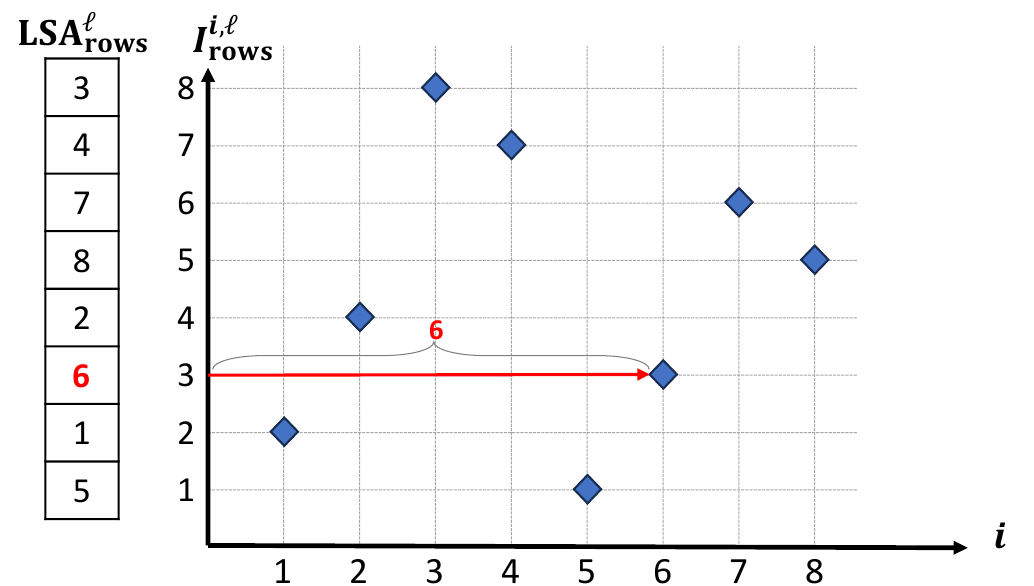} 
    \label{fig:subB}
    \end{center}
 \end{subfigure}\\
 \vspace{2mm}
 \caption{ (a) An example of $\LSA^\ell_{\rows}$. 
   Every cell in $\LSA^\ell_\rows$ contains an index corresponding to a horizontal word in the matrix starting in column $\ell$.
  The (indices representing the) words appear bottom-up in lex-order.\\
  (b) A visualization of the points stored in $D^\ell_\rows$.
  Every point corresponds to a horizontal word. 
  The height of every point corresponds to the location of the corresponding word in $\LSA^\ell_\rows$.
  The horizontal location of a point represents the index of its appearance in the string.
  \label{fig:LSARangeDSExample}}
   \end{center}
   \vspace*{-3mm}
\end{figure}

When looking for matching frames in an $n\times m$ matrix $M$, we make use of the following data structures, which all our algorithms create during their preprocessing phase.
\begin{itemize}
    \item For each column $\ell\in[m]$ we use
    \begin{enumerate}
        \item a lex-sorted array $\LSA_\rows^\ell$ of the strings $\{M[i][\ell..m]\mid i\in[n]\}$ (see \cref{fig:subA});
        \item an $\LCP$ structure $\LCP_\rows^\ell$ over $\LSA_\rows^\ell$;
        \item a range query structure $D_\rows^\ell$, containing all pairs $\{(i, I_\rows^{i,\ell})\mid i\in [n]\}$, where $I_\rows^{i,\ell}$ is the index of the string $M[i][\ell..m]$ in $\LSA_\rows^\ell$ (see \cref{fig:subB}).
    \end{enumerate}
In addition, we build the same three structures for the set of all strings of the form $\overleftarrow{M[i][1..\ell]}$, denoted as $\LSA_{\overleftarrow\rows}^\ell$, $\LCP_{\overleftarrow\rows}^\ell$ and $D_{\overleftarrow\rows}^\ell$.
    \item Symmetrically, for each row $u\in[n]$ we use
    \begin{enumerate}
        \item a lex-sorted array $\LSA_\columns^u$ of the strings $\{M[u.. n][i]\mid i\in[m]\}$;
        \item an $\LCP$ structure $\LCP_\columns^u$ over $\LSA_\columns^u$;
        \item a range query structure $D_\columns^u$, containing all pairs $\{(i, I_\columns^{u,i})\mid i\in[m]\}$, where $I_\columns^{u,i}$ is the index of the string $M[u \ldots n][i]$ in $\LSA_\columns^u$.
    \end{enumerate}
In addition, we build the same three structures for the set of all strings of the form $\overleftarrow{M[1 .. u][i]}$, denoted as $\LSA_{\overleftarrow\columns}^u$, $\LCP_{\overleftarrow\columns}^u$ and $D_{\overleftarrow\columns}^u$.
\end{itemize}

In the remainder of this section we describe an algorithm constructing the data structures for the rows in $\Otild(nm)$ time.
The data structures for the columns can be built similarly.

The algorithm creates the string $M_{\rows} = M[1][1.. m] \cdot \$_1\cdot  M[2][1.. m] \cdot \$_2\cdot \ldots \cdot  M[n][1.. m]\cdot \$_n$, where $\$_1<\cdots <\$_n$ are distinct characters not in $\Sigma$, and build its suffix array $\SA_{\rows}$ using \cref{lem:sa_construct}. 
Then it initializes $\LSA_{\rows}^\ell$ for every $\ell \in [m]$ as an empty array and use $\SA_{\rows}$ to populate these arrays.
Namely, the algorithm scans $\SA_{\rows}$ from left to right.
The suffix starting at position $i$ of $M_\rows$ corresponds to the horizontal substring $M[j][\ell .. m]$ such that $j=\lceil i/(m+1)\rceil$ and $\ell = i \bmod (m+1)$ (unless $\ell = 0$). 
When scanning $M[j][\ell ..m]$,
the algorithms appends the string $M[j][\ell..m]$ to $\LSA_{\rows}^\ell$.

Next, the algorithm constructs $\LCP(M_{\rows})$ using \cref{lem:LCPds}. 
Using this $\LCP$ structure, one can compute any $\LCP$ query within any array $\LSA_{\rows}^\ell$ in constant time.
Indeed, in order to obtain the $\LCP$ of $M[i][\ell .. m]$ and $M[j][\ell.. m]$, one can query $\LCP(M_{\rows})$ with the pair of indices $(i-1)(m+1) + \ell$,  $(j-1)(m+1) + \ell$.

In order to construct $D_{\rows}^{\ell}$, the algorithm views $\LSA_{\rows}^{\ell}$ as a permutation of indices and computes the inverse permutation $\ILSA_{\rows}^{\ell}$.
Then it generates all the points $(i, I_{\rows}^{\ell,i}) = (i, \ILSA_{\rows}^{\ell}[i])$ and builds a $2$-dimensional orthogonal range data structure over these points using \cref{lem:RangeQueries}.

The same data structures for the strings of the form $\overleftarrow{M[i][1..\ell]}$ are obtained by running the same procedures over the string $\overleftarrow{M_{\rows}}$.

\para{Complexity.}
    The time and space complexities for constructing the suffix array and $\LCP$ data structures are $O(n\log n)$ by \cref{lem:sa_construct} and \cref{lem:LCPds}. 
    A permutation can be inverted in $O(n)$ time and space.
    The 2-dimensional orthogonal range data structure over $n$ points can be built in $O(n\log n)$ time and space (\cref{lem:RangeQueries}) for each $\ell \in [m]$.
    Thus, the overall time and space complexity for the described preprocessing is $O(nm\log (nm))$.

\section{The Segment Compatibility Data  Structure}\label{sec:SCDS}
In this section we present the \emph{segment compatibility data structure} ($\SCDS$), which is at the core of our maximum matching frame algorithm (see \cref{subsec:tall}).
We start with technical definitions.

\para{Segment, aligned pair, compatible pairs.}
A horizontal (resp. vertical) \emph{segment} is a triplet $(i,j_{1},j_{2})$ (resp. $(i_{1},i_{2},j)$) with $j_1 < j_2$ (resp. $i_1 < i_2$). It represents the horizontal (resp. vertical) segment in the plane connecting the points $(i,j_1)$ and $(i,j_2)$ (resp. $(i_1,j)$ and $(i_2,j)$).
A pair $(s_1,s_2)$ of horizontal segments is \emph{aligned} if  $s_1 = (i_1,j_1,j_2)$ and $s_2 = (i_2,j_1,j_2)$ for some $i_1<i_2,j_1<j_2 \in \N$.
Such a pair has \emph{distance} $|i_2-i_1|$.
Symmetrically, a pair of vertical segments $(s_1,s_2)$ is aligned if  $s_1 = (i_1,i_2,j_1)$ and $s_2 = (i_1,i_2,j_2)$ for some $i_1<i_2,j_1<j_2\in \N$.
Such a pair has \emph{distance} $|j_2-j_1|$.     

\begin{figure}[ht!]
  \begin{center}
 \includegraphics[scale=0.5]{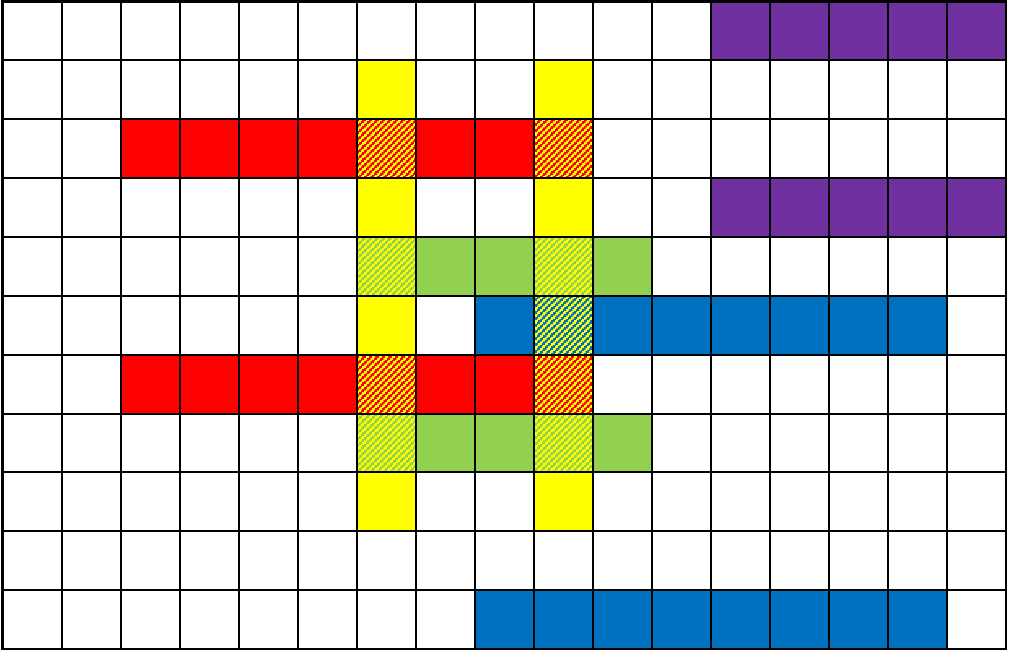} 
  \caption{ An example of horizontal and vertical pairs of aligned segments.
  Every pair of monochromatic lines is an aligned pair of segments.
  The red, green, blue, and purple pairs are horizontal. The yellow pair of vertical segments is compatible with the red and with the green pair.
  \label{fig:AlignedSegments}}
   \end{center}
\end{figure}

An aligned pair of horizontal segments $(i_1,j_1,j_2)$ and $(i_2,j_1,j_2)$ and an aligned pair of vertical segments $(a_1,a_2,b_1)$ and $(a_1,a_2,b_2)$ are \emph{compatible} if and only if $a_1\le i_1 \le i_2 \le a_2$, and $j_1\le b_1 \le b_2 \le j_2$.
The notions of aligned pair and compatible pairs are illustrated by \cref{fig:AlignedSegments}.

The $\SCDS$ stores a set of aligned pairs of vertical segments and supports the query
\begin{itemize}
    \item $\MaxComp(h_1,h_2)$: given an aligned pair $(h_1,h_2)$ of horizontal segments, return a pair $(v_1,v_2)$ with the maximum distance among the stored pairs compatible with $(h_1,h_2)$, or return $\anull$ if no stored pair is compatible with $(h_1,h_2)$.
\end{itemize}

\begin{lemma}
    \label{lem:SCDS}
    Given a set $T$ of $t$ aligned pairs of vertical segments, the $\SCDS$ with $O(\log^3 t)$ query time can be built in $O(t\log^3 t)$ time.
\end{lemma}
\begin{proof}
For each aligned pair $P=\big((a_1,a_2,b_1),(a_1,a_2,b_2)\big)$, we define a 4-dimensional point $\point(P)= (a_1,a_2,b_1,b_2)$ with the value $v(\point(P))=b_2-b_1$.
Then we build, for the set of points $\{\point(P)\mid P\in T\}$, a 4-dimensional range data structure $D$ with $\maximum$ queries.

Let $(h_1,h_2)=\big((i_1,j_1,j_2),(i_2,j_1,j_2)\big)$ be a pair of aligned horizontal segments and let $R=([-\infty,i_1],[i_2,\infty],[j_1,j_2],[j_1,j_2])$.
It is clear that a pair $P$ is compatible with $(h_1,h_2)$ if and only if $\point(P) \in R$. 
Hence, to perform the query {$\MaxComp(h_1,h_2)$}, we query $D$ with $\maximum(R)$ and return the output. 

Due to \cref{lem:RangeQueries}, the construction time and the query time are as required.
\end{proof}

\section{Maximum Matching Frame}

In this section we prove \cref{thm:max_frame}, describing an algorithm with the announced time complexity.
We assume that the input matrix $M$ contains a maximum matching frame $(u,d,\ell,r)$ whose \emph{height} $d-u$ is smaller than or equal to its \emph{width} $r-\ell$.
To cover the complementary case, the algorithm is applied both to the original matrix $M$ and to its transpose $M^\top$ and then the maximum result is reported.

Our algorithm chooses a parameter $x$ and distinguishes between \emph{short} frames of height at most $x$ and \emph{tall} frames with height larger than $x$.
It processes the two types of frames separately and returns the maximum between two solutions.

\subsection{Algorithm for Short Frames}\label{subsec:short}
In this section we prove the following lemma:
\begin{lemma}\label{lem:max_frame_short}
    There is an algorithm that for a given $x\in[n]$ finds, in $\Otild(n \cdot m\cdot x)$ time and $O(n)$ additional space, a maximum matching frame of height at most $x$. 
\end{lemma}

\begin{proof}
    For every two rows $u',d' \in [n]$ such that $d' \in [u'+1..  u'+x]$ the algorithm works as follows.
    First, the algorithm finds all maximal ranges $[a .. b]$ such that $M[u'][a .. b] = M[d'][a .. b]$.
    By ``maximal'' we mean that a range can not be extended to the right or to the left while keeping equality. 
    Note that all maximal ranges are disjoint.
    For $k\in [m]$ we denote the vertical string $M[u'..d'][k]$ by $S_k$. 
    
    Let $[a..b]$ be a maximal range. 
    For every vertical string $S_k$ with $k\in[a..b]$ we find its leftmost and rightmost occurrences in the range $[a..b]$.
    This is achieved by initializing an empty dictionary $D_{a,b}$ and scanning the range $[a..b]$ left to right. 
    For each $k\in[a..b]$ the algorithm computes the fingerprint $f$ in $\LSA_{\columns}^{u'}$ of the string $S_k$ (see \cref{def:fingerprint}). 
    If $f$ is not in $D_{a,b}$, we add $f$ to $D_{a,b}$ and update both the leftmost and rightmost occurrence of $S_k$ to be $k$. 
    If $f$ is already in $D_{a,b}$, we update the rightmost occurrence of $S_k$ to be $k$.

    After completing the scan, the algorithm finds a vertical string $S_k$ such that the distance between the leftmost occurrence $\ell'$ and the rightmost occurrence $r'$ of $S_k$ is maximal. 
    If $\ell'<r'$, we call the frame $(u',d',\ell',r')$ the $(a,b)$-range candidate of $(u',d')$; otherwise, there is no such candidate.
    Among all maximal ranges $[a..b]$, an $(a,b)$-range candidate with the maximal perimeter is the $(u',d')$-candidate (if there are no $(a,b)$-range candidates for $(u',d')$, there is no $(u',d')$ candidate).
    The algorithm outputs a $(u',d')$-candidate with the maximal perimeter over all pairs of rows $(u',d')$ or returns $\anull$ if there are no  such candidates.

\para{Correctness.}
Let $F' = (u',d',\ell',r')$ be the frame returned by the algorithm.
Then $F'$ is the $(a,b)$-range candidate of $(u',d')$ for some range $[a..b]$ such that $a\le\ell'<r'\le b$. Then, the equality $M[u'][a..b]=M[d'][a..b]$ implies $M[u'][\ell'..r']=M[d'][\ell'..r']$, while $M[u'..d'][\ell']=M[u'..d'][r']$ by the choice of $\ell',r'$.
Hence, $F'$ is matching.

Let $F=(u,d,\ell,r)$ be a maximum matching frame among the frames of height at most $x$.
When the algorithm iterates over the rows $u,d$, it identifies a range $[a..b]$ such that $a\le \ell<r\le b$.
Let $\hat{F} = (u,d,\hat{\ell},\hat{r})$ be the $(a,b)$-range candidate of $(u,d)$.
Since $F$ is a valid choice for this candidate, the inequality $r-\ell\le \hat{r}-\hat{\ell}$ holds, implying $\per(F)\le \per(\hat{F})\le \per(F')$.

\para{Complexity.} For a pair of rows $(u',d')$, identifying the maximal ranges takes $O(m)$ time. 
A maximal range $[a..b]$ requires $O(b-a)$ dictionary operations, each taking $O(\log n)$ time using, for example, an AVL tree~\cite{AVL62}. 
Since all the maximal ranges of $(u',d')$ are disjoint, their lengths sum to at most $m$, leading to the running time $\Otild(m)$ for $(u',d')$.

Since $d' \in [u'+1 .. u'+x ]$, there are $O(n\cdot x)$ pairs of rows to process.
Therefore, the total running time of the algorithm is $\Otild(n \cdot m\cdot x)$.
Since the algorithm considers every pair of rows $(u',d')$ separately, the (additional) space usage of the algorithm is $O(n)$.
\end{proof}

\subsection{Algorithm for Tall Frames}\label{subsec:tall}
In this section, we prove the following lemma:
\begin{lemma}\label{lem:max_frame_long}
     There is an algorithm that for a given $x\in[n]$ finds, in $\Otild(\frac{n \cdot m^2}{x})$ time and $\Otild(m^2)$ additional space, a maximum matching frame of height at least $x$.
\end{lemma}

Given a frame $F=(u,d,\ell,r)$ and a position $p=(i,j)$ such that $i\in[u..d]$ and $j\in [\ell..r]$, we say that $p$ is \emph{contained} in $F$ and $F$ \emph{contains}  $p$. 
We say that $F$ is a $(p,H,W)$-frame if $d-u\in[H/2..H]$, $r-\ell\in[W/2..W]$, and $F$ contains $p$.
We introduce an algorithm that finds a maximum matching $(p,H,W)$-frame and use it as a subroutine of the algorithm finding the maximum matching tall frame. 

\begin{lemma}
\label{lem:itai_alg}
     Given a position $(i,j)$ in $M$ and a pair of positive integers $(H,W)\in [n]\times [m]$, there is an algorithm finding a maximum matching $((i,j),H,W)$-frame in $\Otild(H^2{+}W^2)$ time and $\Otild(W^2)$ additional space.
\end{lemma}

\begin{proof}
    For every pair $(\ell,r) \in [m]^2$ such that $r - \ell \in [W/2 .. W]$ and $j \in [\ell.. r]$, the algorithm finds the maximal aligned agreement between the columns $\ell$ and $r$ intersecting the $i$th row by executing two $\LCP$ queries.
    First the algorithm queries $\LCP_\columns^{i}$ to obtain the maximal $d'$ such that  $M[i.. d'][ \ell] = M[i..d'][r]$.
    Similarly, the algorithm queries $\LCP_{\overleftarrow\columns}^{i}$ to obtain the minimal $u'$ such that $M[u' .. i][ \ell] = M[u' .. i][r]$.
    Then the algorithm stores the pair of segments $s_1 = (u',d',\ell)$ and  $s_2 = (u',d',r)$.
    To conclude this part, the algorithm constructs an $\SCDS$ over all stored pairs.
    
    Next, the algorithm iterates over all pairs $(u,d)\in [n]^2$ such that $d - u \in [H/2 .. H]$ and $i \in [u.. d]$.
    For each such pair, the algorithm queries the data structures $\LCP_\rows^{j}$ and $\LCP_{\overleftarrow\rows}^{j}$ (similar to the above computation of vertical agreements), obtaining the minimal $\ell'$ and the maximal $r'$ such that $M[u][\ell' .. r'] = M[d][\ell' .. r']$.
    The algorithm then constructs the horizontal aligned pair of segments $s^h_1 = (u,\ell',r')$ and $s^h_2 = (d,\ell',r')$. 
    The algorithm queries $\SCDS$ for $(s^{v}_{1},s^{v}_{2})\leftarrow\MaxComp(s^h_1,s^h_2)$.
    Let $s^v_1 = (t_1,t_2,\ell)$ and  $s^v_2 = (t_1,t_2,r)$.
    We call the frame $(u,d,\ell,r)$ the \emph{$(u,d)$-optimal frame}. 
    If the query $\MaxComp(s^h_1,s^h_2)$ returns $\anull$, there is no $(u,d)$-optimal frame. 
    The algorithm reports the $(u,d)$-optimal frame with the maximum perimeter among all pairs $(u,d)$, or returns $\anull$ if no such frames were found.

    \para{Correctness.} By construction, each frame $(u,d,\ell,r)$ identified by the algorithm is a $(p,H,W)$-frame. 
    We proceed to show that it is a matching frame.
    Recall that $(u,d,\ell,r)$ was obtained from two compatible pairs of segments $s^v_1$, $s^v_2$ and $s^h_1$, $s^h_2$.
    Notice that for the pair $s^v_1 = (u_v,d_v,\ell)$ and $s^v_2 = (u_v,d_v,r)$ to be compatible with $s^h_1 = (u,\ell_h,r_h), s^h_2 = (d,\ell_h,r_h)$, the inequalities $u_v \le u$ and $d_v \ge d$ must hold.
    By the construction of $s^v_1$ and $s^v_2$ we have $M[u_v .. d_v][\ell] = M[u_v .. d_v][r]$ and then $M[u.. d][\ell] = M[u .. d][r]$.
    In a similar way, one can prove $M[u][\ell .. r] = M[d][\ell .. r]$, showing that $(u,d,\ell,r)$ is a matching frame as required.

    To conclude the correctness of our algorithm, we need to show that some maximum matching $(p,H,W)$-frame is $(u,d)$-optimal for some $(u,d)$.
    Let $(u_t,d_t,\ell_t,r_t)$ be a maximum matching $(p,H,W)$-frame.
    For $(u_t,d_t)$, the algorithm creates the horizontal aligned pair $s^h_1 = (u_t,\ell_h,r_h),s^h_2=(d_t,\ell_h,r_h)$.
    Since $M[u_t][\ell_t .. r_t]=M[d_t][\ell_t .. r_t]$, we have $\ell_h \le \ell_t$ and $r_h \ge r_t$.
    By a similar argument, when constructing the $\SCDS$, the algorithm creates a vertical aligned pair $s^v_1 = (u_v,d_v,\ell_t)$, $s^v_2 = (u_v,d_v,r_t)$ with $u_v \le u_t$ and $d_v \ge d_t$.
    Denote the output of $\MaxComp(s^h_1,s^h_2)$ by $\big((u',\ell',r'),(d',\ell',r')\big)$.
    One has $r' - \ell' \ge r_t - \ell_t$ since the pair $(s^v_1,s^v_2)$ is compatible with $(s^h_1,s^h_2)$.
    Then $(u_t,d_t,\ell',r')$ is a matching frame with perimeter $2(d_t - u_t + r' - \ell') \ge 2(d_t-u_t + r_t - \ell_t)$.
    Due to the maximality of the perimeter of $(u_t,d_t,\ell_t,r_t)$, we have that $(u_t,d_t,\ell',r')$ is a maximum matching $(p,H,W)$-frame.
    
    \para{Complexity.} It can be easily shown that there are $O(W^2)$ pairs $(\ell,r)$ satisfying $r - \ell \le W$ and $j \in [\ell.. r]$.
    Similarly, there are  $O(H^2)$ pairs $(u,d)$ satisfying $d - u \le H$ and $i \in [u.. d]$. 
    By \cref{lem:SCDS}, the construction of the $\SCDS$ takes $\Otild(W^{2})$ time. The algorithm then applies $O(H^{2})$ queries to the $\SCDS$ and the overall complexity is $\Otild(W^2 + H^2)$.
    The additional space usage of the algorithm is dominated by the $\SCDS$ data structure of size $\Otild(W^2)$.
\end{proof}

\begin{proof}[Proof of \cref{lem:max_frame_long}]\label{proof:max_frame_long}
The algorithm iterates over all pairs $H,W \in \{x\cdot 2^k \mid k \ge 1\}$ such that $H\le W< 2m$.
For a pair $(H,W)$, the algorithm runs the subroutine from \cref{lem:itai_alg} for every position $(i,j)\in[n]\times[m]$ such that $i\modulo H/2=0$ and $j\modulo W/2=0$.
Finally, the algorithm reports the maximum matching frame among all outputs of this subroutine.

\para{Correctness.}
Since every instance of the subroutine from \cref{lem:itai_alg} reports a matching frame or a $\anull$, the algorithm also reports a matching frame (or a $\anull$).
Let $F=(u,d,\ell,r)$ be a maximum matching frame of height at least $x$.
Let $W$ (resp. $H$) be the smallest number in $\{x \cdot 2^k \mid k \ge 1 \}$ which is at least $r-\ell$ (resp. $d-u$).
Then there exist $i\in[u..d]$ and $j\in[\ell..r]$ such that $i\modulo H/2=0$ and $j\modulo W/2=0$.
Hence the algorithm ran the subroutine for $((i,j),H,W)$-frames and got reported a matching frame $F'$ with $\per(F') \ge \per(F)$.
Therefore, the algorithm returns a maximum matching frame.

\para{Complexity.}
For a given pair $(H,W)$, the subroutine of \cref{lem:itai_alg} was called for $\floor{\frac{2n}{H}}\cdot\floor{\frac{2m}{W}}$ points $(i,j)$. 
In total, these calls cost $\Otild\big(\frac{nm}{HW} (W^2 + H^2)\big) = \Otild\big(nm \frac{W}{H}\big)$ time.
Therefore, the algorithm runs in $\Otild(nm)\cdot\sum_{H,W} \frac{W}{H}$ time, where the summation is over all possible pairs.
Let $t=\ceil{\log\frac{m}{x}}$.
Since $x \leq H \leq W < 2m$, we have $\sum_{H,W} \frac{W}{H}=2^t+2\cdot 2^{t-1}+3\cdot 2^{t-2}+\cdots\le 4\cdot 2^t = O(\frac{m}{x})$.
The time bound from the lemma now follows.
The additional space usage of the algorithm is dominated by the space of the largest instance of \cref{lem:itai_alg}, which is $\Otild(W^2)$ for some $W$. Since $W<2m$, we have the required bound $\Otild(m^2)$.
\end{proof}

\subsection{Combining the Short and Tall Algorithms}
In this section, we combine the results of \cref{subsec:short} and \cref{subsec:tall} to prove \cref{thm:max_frame}.

\begin{proof}[Proof of \cref{thm:max_frame}]
Applying the algorithm of \cref{lem:max_frame_short} and the algorithm of \cref{lem:max_frame_long} 
with the same threshold $x=\sqrt{m}$ and reporting the maximum frame between both outputs yields an algorithm with running time $\Otild(nm\cdot\sqrt{m})$.
We run the same scheme for the transposed matrix $M^\top$ and $x=\sqrt{n}$, which takes $\Otild(nm \cdot \sqrt{n})$ time.
In total, processing both $M$ and $M^\top$ takes $\Otild(nm\cdot \sqrt{\max \{n,m \} })$ time.
The space usage of the algorithm is dominated by the preprocessed data, which takes $\Otild(nm)$ space.

Notice that $d-u \le r-\ell$ for all considered frames, yielding $d-u \le \min\{n,m\}$.
Therefore, applying \cref{lem:max_frame_short} to both $M$ and $M^\top$ with $x= \min\{n,m\}$ provides an alternative algorithm that outputs the maximum matching frame within $\Otild(nm \cdot \min\{n,m\})$ time. 
Choosing the faster between the two above algorithms implies \cref{thm:max_frame}.
\end{proof}

\section{Approximation Version}\label{sec:approx}
In the $(1-\eps)$-approximation version of the problem, the goal is to find, given a matrix $M$ with a maximum matching frame $F$, a matching frame $F'$ in $M$ with $\per(F')\ge (1-\eps)\per(F)$.
Our algorithm reduces the problem to multiple instances of a decision problem defined below.
The reduction is shown in \cref{lem:reduction} below and the decision problem is solved in \cref{sec:app_dec}.

\para{Decision problem.}
The input for this problem is a matrix $M$, and an \emph{inner rectangle} $(u_\rect,d_\rect,\ell_\rect,r_\rect)$ in $M$. 
A frame $(u,d,\ell,r)$ in $M$ is \emph{surrounding} if $(u_\rect,d_\rect,\ell_\rect,r_\rect)$ is strictly inside it; formally, if $u<u_\rect\le d_\rect< d$ and  $\ell < \ell_\rect \le r_\rect <r$. 
The goal in this version of the problem is to output a surrounding matching frame $(u,d,\ell,r)$ or report that no such frame exists in $M$. 
In \cref{sec:app_dec}, we show that this problem can be solved in near-linear time, by proving the following lemma.
\begin{lemma}\label{lem:decisionsur}
Given an $n\times m$ matrix $M$ with an inner rectangle $(u_\rect,d_\rect,\ell_\rect,r_\rect)$, there is an algorithm that finds, in $\Otild(nm)$ time and space, a surrounding matching frame in $M$ or reports that no such frame exists.
\end{lemma}

Via an application of a $2$-dimensional variant of the so-called \emph{standard trick} \cite{CC07,CKRPRWZ22}, we obtain the following reduction.

\begin{lemma}
\label{lem:reduction}
    Let $a=1+\eps/3$.
    For every $(h,w)\in [\log_an] \times [\log_am]$ such that $a^h,a^w \ge 2$, there is a set $\M_{h,w}$ of sub-matrices, each associated with an inner rectangle, such that the following properties are satisfied:
    \begin{enumerate}
        \item \label{prop:mhw1} $|\M_{h,w}| = O(\frac{nm}{\eps^2 a^{h+w}})$.
        \item \label{prop:mhw2} For every sub-matrix $M' \in \M_{h,w}$, $|M'|=O(a^{h+w})$.
        \item \label{prop:mhw3} For every frame $(u,d,\ell,r)$ with $d-u\in [a^h .. a^{h+1} -1]$ and $r - \ell \in [a^w .. a^{w+1}-1]$ there is a sub-matrix $M' \in \M_{h,w}$ such that $(u,d,\ell,r)$ is a surrounding frame in $M'$ with respect to its inner rectangle.
        \item \label{prop:mhw4} For every surrounding frame $F$ in any $M'\in\M_{h,w}$,  $\per(F)\ge (1-\eps) \big( 2 (a^{w+1}+a^{h+1}) \big)$.
    \end{enumerate}

    The inner rectangles and the corners of the sub-matrices in $\M_{h,w}$ can be obtained in $O(|\M_{h,w}|)$ time and space given $h$ and $w$.
\end{lemma}

\begin{proof}
Fix $(h,w) \in [\log_an ]\times [\log_am ]$.
We define several numeric values that are used repeatedly by our reduction, namely $\delta_w = \floor{\frac{\eps  a^{w+1}}{3}} $, $\delta_h = \floor{\frac{\eps  a^{h+1}}{3}}$, 
$W_w = \ceil{a^{w+2}}$, and $H_h = \ceil{a^{h+2}}$.
For convenience, assume without loss of generality that both $\frac{n- H_h}{\delta_h}$ and $\frac{m- W_w}{\delta_w}$ are integers.
Otherwise, the algorithm adds dummy rows and columns to the right and to the bottom sides of the matrix with distinct unique characters not in $\Sigma$ until $\delta_h$ divides $n- H_h$ and $\delta_w$ divides $m-W_w$.
The set $M_{h,w}$ of sub-matrices of $M$ is defined as follows:
\[
\M_{h,w} = \big\{M[\alpha\delta_h +1.. \alpha\delta_h + H_h][\beta\delta_w +1.. \beta\delta_w + W_w] \mid \alpha \in  [ 0.. \tfrac{n- H_h}{\delta_h} ] \text{ and } \beta \in [0.. \tfrac{m- W_w}{\delta_w} ]\big\}.
\] 
In words, those are all sub-matrices with width $W_w-1$ and height $H_h-1$, having their upper left corner in a cell $(x',y')$ of $M$ such that $x'\bmod\delta_h = y'\bmod\delta_w=1$.
Note that \cref{prop:mhw1,prop:mhw2} are trivially satisfied.
Additionally, it is clear that the corners of each sub-matrix can be obtained in constant time. 

\cref{prop:mhw3} is obtained by combining the following two claims.

\begin{claim}\label{lem:submatrixcontainframe}
    Every frame $(u,d,\ell,r)$ with $d-u \in [a^h .. a^{h+1} - 1]$ and $r-\ell \in [a^w .. a^{w+1} - 1]$ is contained in some $M' \in \M_{h,w}$.
\end{claim}
\begin{claimproof}
    Let $x$ (resp. $y$) be the largest integer multiple of $\delta_h$ (resp. $\delta_w$) that is smaller than $u$ (resp. $\ell$).
    By definition, $\M_{h,w}$ contains a sub-matrix $M' =M[x+1..x+H_h] [y+1..y +W_w]$.
    In order to prove that $(u,d,\ell,r)$ is fully contained inside $M'$, we need to show that 
    (1) $x < u$, (2) $y < \ell$, (3) $x+H_h \ge d$ and (4) $y+W_w \ge r$.
    Conditions (1), (2) are immediate from the choice of $x$ and $y$.
    Let us show (3).
    The choice of $x$ also implies $x + \delta_h \ge u$.
    Therefore, 
    \begin{multline*}
    x + H_h\ge u-\delta_h+H_h=u-\floor{\frac{\eps  a^{h+1}}{3}}+\ceil{a^{h+2}}
    \\
    \ge u- \frac{\eps  a^{h+1}}{3} + a^{h+2} = u + a^{h+1} \left(a-\frac{\eps}{3}\right) = u+ a^{h+1}.
    \end{multline*} 
    By conditions of the lemma, $d - u < a^{h+1}$, so we obtain $x + H_h > d$ as required. 
    Condition (4) can be shown in the same way.
\end{claimproof}
For each sub-matrix $M' =M[x+1..x+H_h] [y+1..y +W_w]$ we define the inner rectangle $R_\rect = (u_\rect,d_\rect,\ell_\rect,r_\rect)=(x+H_h - \ceil{a^h} +1 , x+\ceil{a^h}-1, y+W_w - \ceil{a^w} +1, y+\ceil{a^w} -1 )$. 
As the further argument does not depend on $x,y$, we assume $x=y=0$ for simplicity.

\begin{claim}\label{lem:largesurrounding}
    If $(u,d,\ell,r)$  is  a frame in $M'$ with $r-\ell \in [a^w .. a^{w+1} - 1]$ and $d-u \in [a^h .. a^{h+1} - 1]$, then $(u,d,\ell,r)$  is a surrounding frame.
\end{claim}
\begin{claimproof}
    Since $d\le H_h$ and $d-u \ge a^{h}$, one has $u \le d - a^{h}  \le H_h - a^h < u_\rect $, as required.
    Since $u \ge 1$, one also has $d \ge a^h + 1 > d_\rect$ as required. 
    The inequalities $\ell < \ell_\rect$ and $r > r_\rect$ are proved in the same way, so $(u,d,\ell,r)$ is surrounding by definition.
\end{claimproof}

To prove \cref{prop:mhw4}, we note that the perimeter of a surrounding frame in $M'$ is at least $2((d_\rect-u_\rect+2)+(r_\rect-\ell_\rect+2))$.
We show that $d_\rect-u_\rect+2 \ge (1-\eps) \cdot a^{h+1}$. 
It can be similarly argued that $r_\rect - \ell_\rect+2 \ge (1-\eps) \cdot a^{w+1}$; the two inequalities together yield \cref{prop:mhw4}.
Recall that $u_\rect =  H_h - \ceil{a^h} +1$, $d_\rect = \ceil{a^h}-1$, $H_h=\ceil{a^{h+2}}$.
Then 
    \[d_\rect - u_\rect +2= 
    \ceil{a^h} - 1 - H_{h} + \ceil{a^h}-1 +2 \ge 
    2a^{h} - a^{h+2} = 
    a^{h+1}\big(\tfrac{2}{a} - a\big) \]
It remains to show that $\frac{2}{a} - a \ge 1-\eps$.
Indeed, 
    \[ \frac{2}{a} - a = \frac{2-(1+2\eps/3+\eps^2/9)}{1+\eps/3}= \frac{1-2\eps/3-\eps^2/3 +2\eps^2/9}{1+\eps/3} = 1 - \eps + \frac{2\eps^2/9}{1+\eps/3} > 1- \eps ,\]
as required.
The lemma is proved.
\end{proof}

With \cref{lem:reduction,lem:decisionsur}, we are ready to prove \cref{thm:apx}.

\begin{proof}[Proof of \cref{thm:apx}]
The algorithm first processes frames of height $1$ or width $1$, applying the algorithm of \cref{lem:max_frame_short} with $x=1$ to both $M$ and $M^\top$.
After that, the algorithm proceeds as follows. 
For every pair $(h,w) \in [\log_a n] \times [\log_a m]$ such that $a^w,a^h \ge 2$, it creates the set $\M_{h,w}$ with the corresponding inner rectangles (see \cref{lem:reduction}) and applies \cref{lem:decisionsur} on every $M' \in \M_{h,w}$ with its inner rectangle.
The algorithm returns the maximum frame among the matching frames returned by algorithms of \cref{lem:max_frame_short} and \cref{lem:decisionsur}.
If neither of these two algorithms reported a frame, then a ``no frames'' answer is reported.

\para{Correctness.}
Let $F=(u,d,\ell,r)$ be a maximum matching frame in $M$.
If $d=u+1$ or $r=\ell+1$, then $F$ is found by the algorithm of \cref{lem:max_frame_short}.
Otherwise, consider the pair $(h,w) \in [\log_an] \times [\log_am]$ such that $d-u\in [a^h .. a^{h+1} -1]$ and $r - \ell \in [a^w .. a^{w+1}-1]$.
By \cref{prop:mhw3} of \cref{lem:reduction}, there is a sub-matrix $M'\in \M_{h,w}$ that contains $F$ as a surrounding frame.
The algorithm in \cref{lem:decisionsur} returns a surrounding matching frame $F'$ in $M'$, and by \cref{prop:mhw4} of \cref{lem:reduction}, $\per(F')\ge (1-\eps) \big( 2 (a^{w+1}+a^{h+1}) \big)$.
Since $\per(F)< 2 (a^{w+1}+a^{h+1}) $, the approximation guarantee is fulfilled.

\para{Complexity.}
Given $h$ and $w$, the running time of the algorithm that obtains $\M_{h,w}$ and the suitable $R_\rect$ is $O(|\M_{h,w}|) \subseteq O(nm/\eps^2)$ by \cref{prop:mhw1} of \cref{lem:reduction}.

Due to \cref{prop:mhw1,prop:mhw2} of \cref{lem:reduction}, 
the sum of the sizes of the matrices in $\M_{h,w}$ is $O\left(\frac{nm}{\eps^2}\right)$.
Hence, applying \cref{lem:decisionsur} on all $M'\in\M_{h,w}$ takes $\Otild\left(\frac{nm}{\eps^2}\right)$ time.
Recall that there are $O(\log_{1+\eps}n\cdot \log_{1+\eps}m)=O(\frac1{\eps^2}\log n\cdot \log m)$ values of $h$ and $w$.
Thus, the total running time of the algorithm is $\Otild(\frac{nm}{\eps^4})$.
Each matrix in $\M_{h,w}$ is processed separately.
The space complexity of processing a matrix is $\Otild(a^{h+w}) = \Otild(nm)$.
The space is reused when each matrix is processed, so the overall space complexity of the algorithm is $\Otild(nm)$.
\end{proof}

\subsection{Interesting Pairs and Interesting Triplets}\label{sec:interesting_combi}

\begin{figure}[htb!]
  \begin{center}
 \includegraphics[scale=0.5]{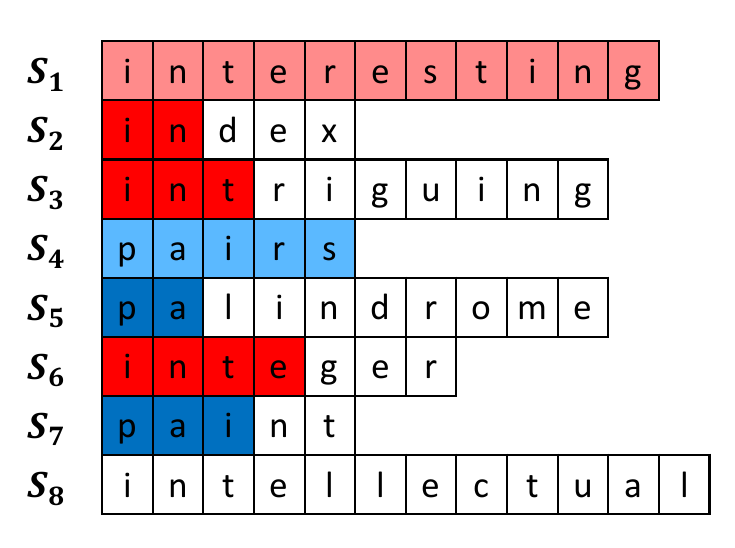} 
  \caption{An example of interesting pairs where the first component of the pair is $S_1$ or $S_4$.
  The rows beginning in red form interesting pairs with $S_1$ and the rows beginning in blue form interesting pairs with $S_4$.
  The color indicates the $\LCP$ of the components of the pair.
  Notice that $(S_1,S_8)$ is not an interesting pair because of $S_6$.
  \label{fig:InterestingPairExample}}
   \end{center}
\end{figure}

In order to prove \cref{lem:decisionsur}, we introduce and study the following notion, illustrated by \cref{fig:InterestingPairExample}.
\begin{definition}\label{def:inter}
    Given a tuple $(S_1,\ldots, S_n)$ of strings, we call a pair $(i,j)$ \emph{interesting} if $i<j$ and for any $\ell$ such that $\ell\in[i+1,j-1]$ one has $\LCP(S_i,S_\ell)<\LCP(S_i,S_j)$.
\end{definition}

Trivially, all pairs of the form $(i,i+1)$ are interesting for any tuple.
The next lemma bounds the number of interesting pairs.

\begin{lemma}\label{lem:inter-count}
    For each $n$-tuple of strings, there are $O(n\log n)$ interesting pairs.
\end{lemma}
\begin{proof}
    For a given tuple $(S_1,\ldots,S_n)$, fix an integer $\ell\in [1..\lceil\log n\rceil]$ and consider the set $\cI_\ell=\{(i,j)\mid (i,j)\text{ is interesting and } j-i \in [2^{\ell-1}.. 2^{\ell}-1]\}$.
    We say that a pair $(i,j)\in \cI_\ell$ is of the \emph{first type} if $i=\max\{i'\mid (i',j)\in \cI_\ell\}$ and of the \emph{second type}  otherwise.
    The following claim is crucial.
\begin{claim}\label{c:n-pairs}
    All pairs of the first type from $\cI_\ell$ have different second components; all pairs of the second type from $\cI_\ell$ have different first components.
\end{claim}
\begin{claimproof}
    The first statement stems directly from the definition of the first type.
    Let us prove the second one.
    Assume by contradiction that $(i,j),(i,j')\in\cI_\ell$ are pairs of the second type, with $j'<j$.
    As $(i,j)$ is not of the first type, $\cI_\ell$ contains a pair $(i',j)$ with $i'>i$.

    We prove the following sequence of inequalities, leading to a contradiction.
\begin{equation*}
    \LCP(S_i,S_{i'}\!)
    \overset{(1)}{<}
    \LCP(S_i,S_{j'}\!)
    \overset{(2)}{=}
    \LCP(S_{j'},S_j)
    \overset{(3)}{=}
    \LCP(S_{i'},S_{j'}\!)
    \overset{(4)}{<}
    \LCP(S_{i'},S_{j})
    \overset{(5)}{=}
    \LCP(S_i,S_{i'}\!),
\end{equation*}   
    Since $2^{\ell-1} \le j-i'$, $2^{\ell-1} \le j'-i$ and $j-i< 2^{\ell} \le j-i'\ +\ j'-i$, we have $i'<j'$.
    Since $(i,j')$ is an interesting pair and $i'\in [i{+}1 .. j'{-}1]$, we obtain (1) by \cref{def:inter}.  
    Since $(i,j)$ is an interesting pair, every $k\in[i{+}1..j{-}1]$ satisfies $\LCP(S_i,S_k)<\LCP(S_i,S_j)$.
    Hence, by \cref{fact:3LCP} we have $\LCP(S_i,S_k)=\LCP(S_k,S_j)$.
    We obtain (2) and (5) by setting $k=j'$ and $k=i'$ respectively.
    Finally, $(i',j)$ is an interesting pair, and $j'\in[i'+1..j-1]$.
    So, \cref{def:inter} gives us (4) and then \cref{fact:3LCP} implies (3).
\end{claimproof}
    \cref{c:n-pairs} says that $\cI_\ell$ contains at most $n$ pairs of the first type and at most $n$ pairs of the second type. As $\ell$ takes $\lceil\log n\rceil$ values, the lemma follows.
\end{proof}

The bound in \cref{lem:inter-count} is asymptotically tight as shown in the following example.
\begin{example}
    Consider the tuple $(S_0,\ldots,S_{n-1})$ of strings over $\{0,1\}$ such that $S_i$ is the binary representation of $i$ written in little endian (the least significant bit first, adding trailing zeroes up to the length $\lceil \log n\rceil$).
    Let $\beta(\ell)$ denote the maximum power of $2$ which divides $\ell$.
    Then $\LCP(S_i,S_j)= \beta(|j-i|)$.
    Hence $(i,j)$ is an interesting pair if and only if $\beta(j-i)$ is strictly greater than $\beta(\ell)$ for all $\ell<j-i$.
    The last condition means exactly that $j-i$ is a power of 2.
    The number of pairs of indices satisfying this condition is $\Omega(n\log n)$, so we have this many interesting pairs.
\end{example}

To relate interesting pairs to our decision problem we need one more notion.
\begin{definition} \label{def:intriples}
    Let $M$ be an $n\times m$-matrix and $\ell\in[m]$.
    A triplet $(u,d,\ell)$ is called \emph{interesting} if the pair $(u,d)$ is interesting for the tuple $(M[1][\ell..m],\ldots, M[n][\ell..m])$.
\end{definition}

\subsection{Finding all interesting triplets}\label{sec:interesting_algo}

\begin{lemma}\label{lem:intersting_efficient}
    All interesting triplets for an $n\times m$ matrix $M$ can be found in $\Otild(nm)$ time. 
\end{lemma}
    We assume that the data structures described in \cref{sec:preproc} are constructed. 
    We process each $\ell\in[m]$ independently, computing all interesting triplets of the form $(u,d,\ell)$. 
    By \cref{def:intriples}, such a triplet is interesting if the pair $(u,d)$ is interesting for the tuple $\S = (S_1,\ldots, S_n)$, where $S_i=M[i][\ell..m]$.
    Below we work with this fixed tuple $\S$.
    The algorithm scans $\S$ string by string; 
    while processing $S_{i}$, the algorithm finds all the interesting pairs $(i,j)$.

    For $i<j\in[n]$, let $L(i,j)$ be the maximum $\LCP$ value between $S_i$ and any $S_k$ for $ k \in[i+1 \ldots j]$. 
    Let $I(i,j) = \min\{k\in [i+1 \ldots j] \mid \LCP(S_i, S_k) = L(i,j) \}$ 
    be the minimum index $k$ with this maximum $\LCP$ value. 
    Using the function $I(i,j)$ we characterize the set of interesting pairs that share the first index $i$.

    \begin{lemma}\label{c:interesting_structure}
    For $i\in[n]$, let $j_1>j_2> \cdots> j_z$ be the second coordinates of all interesting pairs of the form $(i,j)$. Then $j_1 = I(i,n)$ and $j_k = I(i, j_{k-1}-1)$ for every $k \in [2 .. z]$.
    \end{lemma}
\begin{proof}
    First we need to prove that $(i,I(i,n))$ is interesting and that there is no interesting pair $(i,j')$ with $j' > I(i,n)$. 
    By the definitions of $L(i,n)$ and $I(i,n)$, for every $j' < I(i,n)$ we have  $\LCP(S_i, S_{j'}) < L(i,n) = \LCP(S_i,S_{I(i,n)})$, so $(i,I(i,j))$ is interesting. 
    Now consider a pair $(i,j')$ with $j' > I(i,n)$. The same definitions imply $\LCP(S_i, S_{j'})\le L(i,n) =\LCP(S_i,S_{I(i,n)})$, so the pair $(S_i,S_{j'})$ is not interesting and we have $j_1 = I(i,n) $ as required.

    Let $k \in [2 ..z]$ and consider the second statement.
    Similar to the above, we argue that the pair $(i,I(i, j_{k-1}-1))$ is interesting and no pair $(i,j')$ such that $I(i, j_{k-1}-1)<j' < j_{k-1}$ is interesting. Hence $I(i, j_{k-1}-1)$ follows $j_{k-1}$ in the list of second coordinates of interesting pairs of the form $(i,j)$, i.e., $j_k=I(i, j_{k-1}-1)$.
\end{proof}
    
We proceed to show how to compute $I(i,j)$ and $L(i,j)$ efficiently.

\begin{lemma}
\label{lem:lijpolylog}
Given $i$ and $j$, $L(i,j)$ can be computed in $O(\log n)$ time.
\end{lemma}
\begin{proof}
Note that if we lex-sort the tuple $(S_i,\ldots,S_j)$, then the maximum $\LCP$ value with $S_i$ would be reached by one of its neighbors $S_{j_{\lef}}$ and $S_{j_{\righ}}$ in the sorted tuple;
we assume $S_{j_\lef}< S_i<S_{j_\righ}$ (one neighbor may absent).
Thus, $L(i,j)=\max\{\LCP(S_i,S_{j_\lef}),\LCP(S_i,S_{j_\righ})\}$.
The algorithm retrieves $j_\lef$ and $j_\righ$ using range queries on $D^\ell_{\rows}$ as detailed below.

Recall that $I_\rows^{x,\ell}$ denotes the index of $S_x$ in $\LSA^\ell_{\rows}$.
Note that $I^{j_\righ,\ell}_\rows$ is the minimal index satisfying $I_\rows^{x,\ell}>I_\rows^{i,\ell}$ with $x \in [i+1 .. j]$.
Hence, in order to get $j_\righ$ one queries $D^\ell_{\rows}$ for a point $(x,I_{\rows}^{x,\ell})$ in the range $[i+1 .. j] \times [I_\rows^{i,\ell}+1 .. \infty]$ that minimizes $I_{\rows}^{x,\ell}$; the first coordinate of this point is $j_\righ$.
Symmetrically, in order to get $j_\lef$ one queries $D^\ell_{\rows}$ for a point $(x,I_{\rows}^{x,\ell})$ in the range $[i+1 .. j] \times [ 1 .. I_\rows^{i,\ell}- 1]$ that maximizes $I_{\rows}^{x,\ell}$; the first coordinate of this point is $j_\lef$.
After retrieving $j_\righ$ and $j_\lef$, one queries the $\LCP^\ell_\rows$ structure for $\LCP(S_i, S_{j_\righ})$ and $\LCP(S_i,S_{j_\lef})$, and outputs the maximum as $L(i,j)$.

Two range queries take $O(\log n)$ time (\cref{lem:RangeQueries} for $d=2$) while two $\LCP$ queries take $O(1)$ time (\cref{lem:LCPds}). The lemma now follows. 
\end{proof}

\begin{lemma}
\label{lem:iijpolylog}
    Given $i$ and $j$, $I(i,j)$ can be computed in $O(\log n)$ time.
\end{lemma}    

\begin{proof}
    The algorithm starts by applying \cref{lem:lijpolylog} to obtain $L(i,j)$ in $O(\log n)$ time.
    Let $P= S_i[1..L(i,j)]$ be the prefix of length $L(i,j)$ of $S_i$. 
    Recall that by definition, $I(i,j)$ is the minimal index $k\in[i+1..j]$ such that $S_k[1.. L(i,j)] = P$.
    Using \cref{lem:sa_fingerprint}, the algorithm finds, in $O(\log n)$ time, a pair of indices $i_P,j_P$ such that $S_z[1.. L(i,j)] = P$ if and only if $I_{\rows}^{z,\ell}\in[i_P..j_P]$.
    After that, the algorithm retrieves $I(i,j)$ by querying $D^\ell_{\rows}$ for the point $(k , I_{\rows}^{k,\ell})$ in the range $[i+1 .. j]\times [i_P .. j_P]$ with the minimal first coordinate.
    This coordinate $k$ is then reported as $I(i,j)$.
    As this query takes $O(\log n)$ time by \cref{lem:RangeQueries} for $d=2$, the lemma follows.
\end{proof}

We are now ready to present the algorithm proving \cref{lem:intersting_efficient}.

\begin{proof}[Proof of \cref{lem:intersting_efficient}]
Let $\ell$ be fixed and $\S=\{S_1,\ldots,S_n\}$ be defined as above.
For each $S_i$, the algorithm finds $j_1 = I(i,n)$ using \cref{lem:iijpolylog}, reports $(i, j_1)$ as an interesting pair (see \cref{c:interesting_structure}), and then  iterate. 
As long as $j_k \neq i+1$, the algorithm finds $j_{k+1} = I(i, j_k-1)$ using \cref{lem:iijpolylog} and reports the interesting pair $(i,j_{k+1})$.
Note that the algorithm is guaranteed to finish the iteration, since the pair $(i,i+1)$ is interesting.

The algorithm spends $O(\log n)$ time per interesting pair by \cref{lem:intersting_efficient};
the number of such pairs is $O(n\log n)$ by \cref{lem:inter-count}.
Multiplying this by $m$ choices for $\ell$, we obtain the required time bound $\Otild(nm)$.
\end{proof}

\subsection{Algorithm for the Decision Variant}\label{sec:app_dec}

In this section we prove \cref{lem:decisionsur}, presenting the required algorithm.

The algorithm starts by modifying $M$ as follows. 
For every $(i,j)\in [u_\rect\ldots d_\rect] \times [\ell_\rect\ldots r_\rect]$, we set $M[i][j] = \$_{i,j}$ with $\$_{i,j}$ being a unique symbol not in $\Sigma$.
Since neither of the changed symbols belongs to a marginal row/column of a surrounding frame, this modification preserves surrounding matching frames.
The following claim clarifies the role of interesting triplets.

\begin{lemma}\label{lem:framethenintressur}
If a matrix $M$ with an inner rectangle $(u_\rect,d_\rect,\ell_\rect,r_\rect)$ contains a surrounding matching frame $(u,d,\ell,r)$, then it contains a surrounding matching frame $(u',d',\ell,r)$ such that $(u',d',\ell)$ is an interesting triplet.
\end{lemma}
\begin{proof}
Let $(u,d,\ell,r)$ be a surrounding matching frame in $M$.
We denote $S_h = M[u][\ell .. r] = M[d] [\ell ..r]$.
Let $u'$ be the maximal index in $[u .. u_\rect - 1]$ such that $M[u'][\ell .. r] = S_h$ and let $d'$ be the minimal index in $[d_\rect+1 .. d]$ such that $M[d'][\ell .. r]= S_h$. 
The frame $(u',d',\ell,r)$ is surrounding by definition and matching by construction (note that $M[u.. d][\ell] = M[u .. d][r]$ implies $M[u'.. d'][\ell] = M[u' .. d'][r]$).
Finally, for arbitrary $d''\in[u'+1..d'-1]$ one has $M[d''][\ell ..r] \ne S_h$. 
If $d''<u_\rect$ or $d''> d_\rect$, this condition holds by the choice of $u'$ and $d'$ respectively. 
Otherwise the condition is guaranteed by uniqueness of the symbols of the inner rectangle.
Hence 
$\LCP(M[u'][\ell ..m], M[d''][\ell ..m])< |S_h|\le \LCP(M[u'][\ell ..m], M[d'][\ell ..m])$,
and the triplet $(u',d',\ell)$ is interesting by definition.
\end{proof}
    
\para{The Algorithm.}
After setting $M[i][j] =\$_{i,j}$ for each $(i,j)\in [u_\rect\ldots d_\rect] \times [\ell_\rect\ldots r_\rect]$, the algorithm  applies the preprocessing described in \cref{sec:preproc} and finds all interesting triplets in $O(nm\log^2 n)$ time by applying \cref{lem:intersting_efficient}.
The final ingredient we need 
is a mechanism verifying, given an interesting triplet $(u,d,\ell)$, if there is a surrounding matching frame $(u,d,\ell,r)$. 
For this purpose, we present the following lemma.
\begin{lemma}\label{lem:verifysurrounding}
    There is an algorithm that, given an interesting triplet $(u,d,\ell)$ of $M$, outputs an integer $r$ such that $(u,d,\ell,r)$ is a surrounding matching frame or reports $\anull$ if no such $r$ exists. 
    The algorithm runs in $O(\log n)$ time.
\end{lemma}
\begin{proof}
    The algorithm reports $\anull$ if $u \ge u_\rect$, or $d \le d_{\rect}$, or $\ell \ge \ell_\rect$.
    Otherwise, it seeks for a value $r$ such that (i) $r \ge r_\rect + 1$, (ii) $M[u][\ell .. r] = M[d][\ell .. r]$, and (iii)
        $M[u..d][r] = M[u..d][ \ell ]$.

    The algorithm queries $\LCP^\ell_\rows$ for $L_{u,d} = \LCP(M[u][\ell .. m] , M[d][\ell .. m])$.
    By definition of $\LCP$, we have $M[u][\ell .. r] = M[d][\ell .. r]$ if and only if $r \le  \ell + L_{u,d} - 1$.
    Hence, conditions (i) and (ii) are satisfied   if and only if $r \in [r_\rect +1 \ldots \ell + L_{u,d}-1] $.
    To check (iii), let $S_v =  M[u..d][\ell]$.
    Using \cref{lem:sa_fingerprint}, the algorithm finds the pair of indices $i_v, j_v$ such that $M[u.. d][r] = S_v$ if and only if $r \in \LSA^u_{\columns}[i_v  .. j_v]$.
    Now the algorithm checks the existence of a value $r$ satisfying (i)--(iii) by querying $D^u_\columns$ for a point within the range $ [r_\rect +1 .. \ell + L_{u,d}-1] \times [i_v .. j_v]$.
    If the queried structure returns a point $(r, I^{u,r}_{\columns})$, the algorithm outputs $r$; otherwise, it reports $\anull$, as there is no value of $r$ such that $(u,d,\ell,r)$ is a surrounding matching frame.
    
    The algorithm performs a single $\LCP$ query ($O(1)$ time by \cref{lem:LCPds}), finds $i_v$ and $j_v$ ($O(\log n)$ time by \cref{lem:sa_fingerprint}), queries $D^u_\columns$ ($O(\log n)$ time by \cref{lem:RangeQueries}), and compares a constant number of integers.
    The lemma follows.
\end{proof}

We are finally ready to prove \cref{lem:decisionsur}.
\begin{proof}[Proof of \cref{lem:decisionsur}]
After finding all interesting triplets, the algorithm applies the subroutine from \cref{lem:verifysurrounding} to every interesting triplet $(u,d,\ell)$. 
If this subroutine outputs $r$, the algorithm outputs the surrounding matching frame $(u,d,\ell,r)$.
If the subroutine outputs $\anull$ for all interesting triplets, then, relying on \cref{lem:framethenintressur}, the algorithm reports that no surrounding matching frame exists.

The algorithm spends $O(nm\log^2(nm))$ for each of three tasks it performs: preprocessing (\cref{sec:preproc}), finding interesting triplets (\cref{lem:intersting_efficient}), and verifying  interesting triplets (\cref{lem:inter-count} and \cref{lem:verifysurrounding}).
Thus, its time (and therefore, space) complexity is $\Otild(nm)$, as required.
\end{proof}

\bibliography{bib.bib}

 \end{document}